\documentstyle[12pt]{article}
\newcommand{\be}{\begin{equation}}
\newcommand{\ee}{\end{equation}}

\newcommand{\el}{\varepsilon}

\begin{document}

\begin{titlepage}

\center{{\bf Q-DEFORMED SOLITONS AND QUANTUM SOLITONS OF THE MAXWELL-BLOCH LATTICE}}

\vspace{1 cm}
{\center{\bf Andrei Rybin${}^{\dag}$, Jussi Timonen${}^{\dag}$, Gennadii Varzugin${}^{\ast}$\\
 and \\
Robin K. Bullough${}^{ \ddag}$}}

\begin{center}{\dag Department of Physics, University of Jyv{\"a}skyl{\"a}}\\
 {PO Box 35, FIN-40351}\\
{Jyv{\"a}skyl{\"a}\\
 Finland}
\end{center}

\begin{center}{$\ast$ Institute of Physics}\\
 {St. Petersburg State University}\\
 {198904, St. Petersburg\\
Russia}
\end{center}
 \begin{center}
{\ddag Department of Mathematics}\\
 {UMIST}\\
 {PO Box 88}
{Manchester M60 1QD}\\
{UK}
\end{center}

\begin{abstract}

\baselineskip 0.5 cm
 We report for the first time exact solutions of a completely integrable nonlinear lattice system for which
the dynamical variables satisfy a $q$-deformed Lie algebra - the Lie-Poisson algebra $su_q(2)$.
The system considered  is a $q$-deformed lattice for which in continuum limit the equations of motion become the
envelope Maxwell-Bloch  (or SIT) equations  describing the resonant
interaction of light with a nonlinear dielectric. Thus the
$N$-soliton solutions we here report are the natural $q$-deformations, necessary for a lattice, of the well-known
multi-soliton and breather solutions of self-induced transparency (SIT). The method we use to find these solutions is a
generalization of the Darboux-B\"acklund dressing method. The extension of these results to quantum solitons is sketched.

\end{abstract}



For communications: Andrei.Rybin@phys.jyu.fi

\end{titlepage}

\section{Introduction}

The Maxwell-Bloch (MB) system of equations have been fundamental to much of theoretical quantum optics and nonlinear optics
since they were first introduced in the late 1960's (some history is in \cite{ryb1} and also in \cite{bul1}).
These MB systems are of abiding theoretical interest. A feature is that their complete integrability is
handed down from the 'reduced MB' or (RMB) equations to the envelope MB (or self-induced transparency (SIT)) equations,
thence, at resonance, to the Sine-Gordon equation (cf. e.g. \cite{bul2,bul3}).
Each member of this hierarchy has important physical applications while the physics of SIT in particular remains a very active
field of current research \cite{bul4}, even into the femto-second pulse regime\cite{ryb1,bul5}.

Our recent paper  \cite{ryb1} followed up  ideas of quantum groups
and their relevance to integrable systems theory and derived a
$q$-deformed lattice version of the envelope MB system together
with its zero-curvature representation: in continuum limit these
lattice equations become the resonant envelope MB (or SIT)
equations. In this paper we now report exact $N$-soliton solutions
of this $q$-deformed dynamical system. Solitons of the lattice
equations were promised in  \cite{ryb1}, and a Riemann-Hilbert
method of solution sketched. But for the pure $N$-soliton solution
reported in this paper it is more convenient to use a variant of
the Darboux-B\"acklund dressing method which (see below) makes an
ansatz for the dressing in terms of appropriate $N$ bound states
eigenvalues.

Historically \cite{bul2,bul6} multi-soliton solutions of the SIT equations were found by the method which become Hirota's method
\cite{bul1}; Lamb~\cite{lam} gave the inverse scattering solutions; the inverse method for the RMB equations was in \cite{bul7} and
\cite{bul2} obtains the multi-soliton solutions for the SIT equations from it; \cite{abl} gave a further account of inverse
scattering for these SIT equations. These several results on inverse scattering confirmed the generality of a method first devised
to solve the Korteweg-de Vries equation \cite{bul1}.

Expressed in terms of the complex slowly varying envelopes
 for  the electric field and polarisation $\varepsilon$, $\rho$ and the real inversion $\cal N$,
the SIT equations can be put in the
form (e.g. \cite{abl}).

 \begin{equation}\begin{array}{ccc}\partial_\xi\el=\langle\rho\rangle,\\
\partial_\tau\rho+2i\eta\rho={\cal N}
\varepsilon,\\
\partial_\tau{\cal N}=-{1\over2}(\varepsilon^\ast\rho+\varepsilon\rho^\ast).
\end{array}\label{mb}\end{equation}

Propagation is along a coordinate $z$  and
$\xi=\Omega x$ with $x=z\slash c$. The time $\tau$ is a retarded time,  $\tau=\Omega(t- x)$;
$\eta=(\omega-\omega_{\scriptscriptstyle 0})\slash2\Omega$ is the detuning, and
$\Omega=2\pi n_{\scriptscriptstyle 0}\omega_{\scriptscriptstyle
0}\mu^2\slash\hbar$ is the coupling constant. The number $n_{\scriptscriptstyle 0}$ is the density of
 2-level atoms with the non-degenerate transition frequency $\omega_0$ ($\mbox{rad}\cdot\mbox{sec}^{-1}$);  $\mu$  is the
 matrix element for dipole allowed transitions at $\omega_0$. The star denotes complex conjugation and $\langle\,.\,\rangle
=\int_{-\infty}^\infty h(\eta)(\,.\,)\,d\eta$ is the average over inhomogeneous
broadening: $ h(\eta)$ is a $\delta$-function in the sharp-line limit case \cite{bul2}.

In \cite{ryb1} we constructed the completely integrable lattice system whose
equations of motion for three dynamical variables $s_n$, $H_n$ and $\beta_n$ at each lattice site $n$ can be written

\begin{eqnarray}
\partial_t\beta_n&=&-\frac{1}{2}q^{2(N_n+H_n)}
(\beta_{n+1}+\beta_n)-\frac{i}{2}q^{2N_n}(s_n+s_{n-1})\nonumber\\
\partial_t s_n &=&-\frac{i}{2}(\beta_n+\beta_{n+1})(1+2\gamma s_n s^\ast_n)
+\frac{1}{2}q^{2(N_n+H_n)}(s_n+s_{n-1})\label{lat}\\
\partial_t H_n &=& \frac{i}{2}(s_n-iq^{2H_n}\beta_n)(\beta_n^\ast +
\beta_{n+1}^\ast) - \frac{i}{2}(s_n^\ast+iq^{2H_n}\beta^\ast_n)(\beta_n  +
\beta_{n+1}).\nonumber
\end{eqnarray}
Here $q^{2N_n}=1+2\gamma \beta_n^\ast\beta_n$, $q=e^\gamma$ and $\gamma>0$, is a real
parameter (a coupling constant -- see below). Reference to  \cite{ryb1} shows that in Eqs.~(\ref{lat})
we use $s_n=\sqrt{2\gamma}\chi_n+iq^{2H_n}\beta_n$: in  \cite{ryb1} the second equation is for $\partial_t\chi_n$.
As can be checked (and cf. \cite{ryb1})  when the lattice spacing
$\Delta\to 0$ for a continuum limit with
\begin{equation}
 \begin{array}{c}
t\to t\Delta^{-1},\, x=n\Delta,\,  \beta_n=\sqrt{\Delta}{\cal E}(x),\,\chi_n=\Delta
S(x),\,H_n=\Delta S^3(x)\\
\gamma=\kappa\Delta\slash2, \,\kappa>0.\end{array}
\label{22}\end{equation}

one reaches the resonant sharp-line form of the envelope  MB (or SIT) equations
 Eqs~(\ref{mb}) via the definitions

\begin{equation}
\varepsilon(\xi,\tau)=2{\cal E}(x,t),\,\rho(\xi,\tau)=-2iS(x,t),
\,N(\xi,\tau)=2S^3(x,t),\label{cont}
\end{equation} with $\Omega=\sqrt{\kappa}$.
Our use of 'lattice Maxwell-Bloch system(LMB) equations' for Eqs.~(\ref{lat}) stems from this fact.

A Hamiltonian  for this LMB system  is \cite{ryb1}

\begin{eqnarray}
{\cal H}^L&=&\frac{1}{2}\sum_{n=1}^M\left \lbrace\sqrt{2\gamma}
\left [\chi^\ast_n
(\beta_{n+1}+\beta_n)+\chi_n(\beta^{\ast}_{n+1}+\beta^{\ast}_n)\right]\right. \nonumber\\
{}&+&
\left. iq^{2H_n}\left(\beta^{\ast}_{n+1}\beta_n-\beta_{n+1}\beta^{\ast}_n\right)
\right\rbrace.
\label{14}\end{eqnarray}

For $M<\infty$ it would be natural to impose periodic boundary conditions. But we shall look for lattice soliton solutions and
here think of $M\to\infty$ with suitable boundary conditions still to be specified. The Poisson brackets of Eq.~(\ref{14}) are

\begin{equation} \{ X_n^\ast,X_m\}=i\{ 2H_n\} \delta_{mn}, \quad
\{ H_n,X_m\}= -i X_n \delta_{mn},\label{sp}\end{equation}
and the quantities $X_n^\ast$, $X_n$ and $H_n$ form    the $su_q(2)$ Lie-Poisson algebra for by $\{\cdot\}$  we mean $\{ x \}=
({q^x -q^{-x}})\slash({2\gamma})$.
This algebra has a central element $ X_nX_n^\ast +\{ H_n\}^2=\{S\}^2$. The variables $X_n^\ast$, $X_n$ enter (\ref{lat})
via $\chi_n=q^{H_n}X_n$ \cite{ryb1}. The variables
$\beta_n$,
$\beta_n^\ast$ (the 'electric fields', see Eqs.~(\ref{22}),(\ref{cont}) above)  satisfy the Lie-Poisson $q$-boson algebra

\begin{equation} \{ \beta_n, \beta_m^{\ast}\}= iq^{2N_n}\delta_{mn},\,
\{ N_n, \beta_m\}= -i\beta_n\delta_{mn}.\label{bos}
\end{equation}

\section{The $q$-deformed solitons}

In~\cite{ryb1} we   obtained the zero-curvature representation of
the system~(\ref{lat}) which means that we constructed an
over-determined linear system for a matrix-function
$\Psi_n(\zeta,t)$ such that

\begin{eqnarray}
\Psi_{n+1}&=&  L (\zeta|n) \Psi_n\label{16a}\\
\partial_t{\Psi_n}&=&  V (\zeta|n) \Psi_n,\label{16b}
\end{eqnarray}
where

\begin{equation}
 V (\zeta|n)=\sum_{j=-2}^2 \zeta^jV_j(n),\,
 L (\zeta|n)=\frac{q^{-N_n-H_n}}{2\gamma}\sum_{j=-2}^2 \zeta^jL_j(n).\label{mat1}
\end{equation}
Here
\begin{equation}
V_0(n)= {2}{i\gamma}(\beta_n s^\ast_{n-1}+ \beta_n^\ast s_{n-1})\sigma^z,
 \, V_{\pm 2}=\mp\frac{1}{4}\sigma^z, \end{equation}
\begin{equation}
V_{+1}(n)=-\frac{\sqrt{2\gamma}}2\left( \begin{array}{cc}0&i\beta_n^\ast\\
s_{n-1}&0\end{array}\right),\,V_{-1}(n)=\frac{\sqrt{2\gamma}}{2}\left(\begin{array}{cc}0&s_{n-1}^\ast\\
-i\beta_n&0\end{array}\right), \label{mat3}
 \end{equation}
while
\begin{equation}
L_0(n)=2i\gamma\left(\begin{array}{cc}\beta_n s_n^\ast&0\\
0&\beta_n^\ast s_n
\end{array}\right) - q^{2(N_n+H_n)}\sigma^z,\, L_{\pm 2}= \frac{1}{2}( \sigma^z \pm I),\label{mat4}
\end{equation}

\begin{equation}
L_{+1}(n)=\sqrt{2\gamma}\left(\begin{array}{cc}0&i\beta_n^\ast\\
s_n&0\end{array}\right),\,L_{-1}(n)=\sqrt{2\gamma}\left(\begin{array}{cc}0&s_n^\ast\\
-i\beta_n&0\end{array}\right). \label{mat5}
\end{equation}
The parameter $\zeta\in {\bf C}$  which appears in Eqs.~(\ref{16a})-(\ref{mat1}) will be thought of  as the {\em spectral parameter}
while in continuum limit (\ref{16b}) is a spectral problem in $L$ in the usual $2\times2$ sense (Zakharov-Shabat linear system~\cite{bul1});
$\sigma^{x,y,z}$ are the Pauli matrices.
The compatibility condition of the two linear systems Eqs.~(\ref{16a}),(\ref{16b}) under isospectral condition $\partial_t\zeta=0$ is
\begin{equation} \partial_t   L (\zeta|n) +    L (\zeta|n)
V (\zeta|n)  -     V  (\zeta| n+1)
  L (\zeta|n) =0\label{zer}\end{equation}
and this coincides with Eqs.~(\ref{lat}), independent of $\zeta$. However, $\zeta=e^{i\gamma\lambda}$, $\lambda\in{\bf C}$ as it
was introduced in \cite{ryb1}; $\lambda$ is a second 'spectral parameter' and the real axis in the $\lambda$-plane is the circle of
the unit radius in the $\zeta$-plane; $\lambda$ is the usual spectral parameter for the equations Eqs.~(\ref{mb}) derived in
continuum limit. Notice that time $t$ is suppressed in Eqs.~(\ref{16a}),(\ref{16b}): an explicit time dependence will be indicated
only where and when it is needed. Reference to Eqs.~(\ref{16a}) and (\ref{16b}) may make plain that the
function
$\Psi_n(\zeta)$ possesses essential singularities of the rank 2 at $\zeta=0,\infty$.  It is also  important to notice that the
linear equations Eqs.~(\ref{16a}) and (\ref{16b}) are  invariant under the transformations
\begin{equation}
\Psi_n(\zeta)\to (-1)^{n-1} \sigma^y\Psi_n^\ast\left(\frac{1}{\zeta^\ast}\right)\sigma^y,\, \Psi_n(\zeta)\to  \sigma^z \Psi_n(-\zeta)\sigma^z
\label{red}
\end{equation}

We can now turn  to the derivation of exact solutions of the LMB system
Eqs.~(\ref{lat}). For  this, as mentioned,  we develop a variant of the  Darboux-B\"acklund dressing
procedure\cite{neu} rather then any inverse scattering
 method~\cite{bul1, fad}. The essence of the dressing procedure
is to choose a 'seed' solution of the system
Eqs.~(\ref{lat}), typically some trivial solution, and construct
from it a new solution associated with additional points $\zeta_\nu$, $\nu=1,\ldots,N$ (say) of the discrete spectrum: thus
$\det\Psi_n(\zeta_\nu,t)=0$ \cite{bul1,neu,jim} for the new solution $\Psi_n(\zeta,t)$.

For initial and boundary conditions observe that for SIT and envelope MB system Eqs.~(\ref{mb}),
 typical experimental situation  is the half-space problem:
an  initial optical pulse enters, supposedly without reflection from $x<0$  into the resonant medium
 $x\ge0$ and here  breaks up into background radiation and a sequence of soliton pulses.
The corresponding mathematical problem is the Cauchy problem at the point $x=0$:
$\varepsilon(x,t)\vert_{x=0}=\varepsilon_0(t)$ together with the  asymptotic boundary conditions (in $t$) that for $x>0$
${\cal N}\to{\cal N}_-$, $\rho\to 0$ as $t\to-\infty$. For the so-called 'attenuator' $N_-$ is the ground state $N_-=-1$ of the
inversion density. For the lattice problem we therefore take  the half-space problem in which
$\beta_n(t)$ and $s_n(t)$  are sufficiently decreasing for $|t|\to\infty$,
while $H_n(t)\to H$ such that  $H$ corresponds to $N_-$. In this way we would look for a solution  in the
half-space $n > 0$, for which it becomes  the Cauchy
problem specified by the conditions

\begin{equation} \beta_n(t)|_{n=1}=\beta_1(t),\,s_n(t)|_{n=1}=s_1(t),\,
 H_n(t)|_{n=1}=H_1(t).\label{1s}\end{equation}

With this as motivation we report in this paper exact $N$-soliton
solutions  derived by the dressing procedure based on the seed
solution
\begin{equation}
\beta_n=0,\,s_n=0,\,H_n=H.\label{seed}
\end{equation}
  the corresponding solution of the linear system Eqs~(\ref{16a}),(\ref{16b}) is then
\begin{equation}
\Psi^{(0)}_n(\zeta,t)=\exp{\left\{-\frac{\sigma^zt}{4}\right\}}
\left(\zeta^2-\frac{1}{\zeta^2}\right)
\left(\begin{array}{cc}z^{n}(\zeta)&0\\
0&\left(-z\left(\frac{1}{\zeta}\right)\right)^{n}\end{array}\right),
\end{equation}
where $z(\zeta)=\frac{1}{2\gamma}\left(\zeta^2 q^{-H}-q^H\right)$, while the
corresponding operator $V^{(0)}(\zeta|n,t)$ has
$V^{(0)}_0=V^{(0)}_{\pm 1}=0$, and
$V^{(0)}_{\pm 2}=\mp\frac{1}{4}\sigma^z$.
 For the $N$-soliton solution of Eqs.~(\ref{lat})
we construct the new solution
 $\Psi^{(N)}_n(\zeta)$ of Eqs.~(\ref{16a}),(\ref{16b})
 through the ansatz

\begin{equation}
\Psi^{(N)}_n(\zeta)=F(\zeta)\Psi_n^{(0)}(\zeta)\label{backl}.
\end{equation}
The function   $F(\zeta)$  is to have  poles only at
 the essential singularities  of $\Psi_n(\zeta)$. As was indicated above these points  are $0$ and $\infty$.
This suggests  the  ansatz
\begin{equation}
F(\zeta,n,t)=F_0(n,t) +\sum_{i=1}^N \zeta^i F_{+i}(n,t) +{\zeta}^{-i}F_{-i}(n,t).\label{backl1}
 \end{equation}
It is convenient to impose the additional conditions on $F(\zeta)$ that
\begin{equation}
\sigma^y{ F}^\ast\left( \frac{1}{\zeta^\ast}\right)\sigma^y=F(\zeta),\, \sigma^z F(-\zeta)\sigma^z=(-1)^N F(\zeta)\label{red1}
\end{equation}
which are obviously compatible with the transformation Eq.~(\ref{red})
We can also normalize the matrix  $F(\zeta)$ so  that for each $(n,t)$

\begin{equation}
F_{-N}={\cal Q}\left(\begin{array}{cc} f(n,t)&0\\0&f^{-1}(n,t) \end{array} \right),\,
F_{N}={\cal Q}^\ast \left(\begin{array}{cc} f^{-1}(n,t)&0\\0&f(n,t) \end{array} \right)_,\label{norm1}
\end{equation}
where the constant ${\cal Q}$ is  independent of $n$ and $t$ and $f(n,t)$ is a real function.
We now choose a set of $N$ points $\left\lbrace \zeta_\nu \right\rbrace_{\nu=1}^N$ where
$\det{\Psi^{(N)}_n(\zeta)}$ is to vanish.
This means

\begin{equation}
F(\zeta_\nu)\Phi(\zeta_\nu)=F\left(\frac{1}{\zeta_\nu^\ast}\right)\sigma^y\Phi^\ast(\zeta_\nu)=0,\label{syst}
\end{equation}
where
$$
\Phi(\zeta_\nu)=\left(\begin{array}{c}\Phi_1(\zeta_\nu)\\
\Phi_2(\zeta_\nu)\end{array}\right)=\Psi_n^{(0)}(\zeta_\nu)\left( \begin{array}{c}1\\-c_\nu \end{array} \right)_,
$$
and $c_\nu $ are constants independent on $n$ and $t$. The set $\lbrace \zeta_\nu,c_\nu \rbrace_{\nu=1}^N$ together constitute a
necessary and complete set of parameters (spectral data) for an $N$-soliton solution~\cite{bul1}.

The system of equations Eqs.~(\ref{syst}) has a unique solution  satisfying conditions
Eqs.~(\ref{red1}),(\ref{norm1})  if we choose

\begin{equation}
{\cal Q}=e^{ i\frac{\pi}{2}N+i\sum_{\nu=1}^N \alpha_\nu},\,\,\, \zeta_\nu=e^{\gamma_\nu+i\alpha_\nu},
\end{equation} where $\gamma_\nu,\, \alpha_\nu\in {\bf R}.$
In so far as $\zeta=e^{i\gamma\lambda}=e^{\gamma_\nu+i\alpha_\nu}$  and $\lambda$ is the spectral parameter for Eqs.~(\ref{mb}) we
are interested in zeros $\zeta_\nu$ defined by the half $\lambda$-plane $\mbox{Im}      \lambda\ge 0$ which lie inside the circle
$|\zeta|=1$ in the $\zeta$-plane.
 The linear system Eqs.~(\ref{16a}),(\ref{16b})  is invariant under the gauge transformation Eqs.~(\ref{backl})
with the potentials written as
\begin{eqnarray}
F_{-N+1}F_{-N}^{-1}&=&\sqrt{2\gamma}\left(
\begin{array}{cc}0&-s_{n-1}^\ast\\-i\beta_n&0
\end{array} \right)\nonumber\\
q^H\frac{f(n+1,t)}{f(n,t)}&=&q^{N_n+H_n}\label{dress_x}
\end{eqnarray}

We turn next to a determination of the matrices $F_{\pm i}(n,t)$.
The conditions Eqs.~(\ref{red1}) suggest that we should take the matrices $F_{-N+2k}$  diagonal, and the matrices
$F_{-N+2k-1}$  off-diagonal in agreement with the first of Eqs.~(\ref{dress_x}).

So will  $F_{-N}^{-1}$ be diagonal from Eq.~(\ref{norm1}) we set
\begin{equation}
F_{-N}^{-1}F_{-N+2k-1}=\left( \begin{array}{cc} 0&y_k\\\tilde y_k&0\end{array} \right),\,
F_{-N}^{-1}F_{-N+2k}=\left( \begin{array}{cc} x_k&0\\0&\tilde x_k \end{array} \right)
\end{equation}
in which  $y_k$, $\tilde y_k$, $x_k$, $\tilde x_k$ are (so far)  arbitrary independent complex numbers.

Then the conditions for the zeros $\zeta_\nu$ Eqs.~(\ref{syst}) mean that we can instead solve

\begin{equation}
(1,0)+\sum_{k=1}^N z_k \sigma_\nu^{2k}=0, \,\nu=1,..,N;\label{28}
\end{equation}
in which the $z_k$ are row-vectors
 $z_k=(x_k,y_k)$, and the matrices $\sigma_\nu=S_\nu\Lambda_\nu S_\nu^{-1}$ in which
  $\Lambda_\nu=\mbox{diag}(\zeta_\nu,1/\zeta_\nu^\ast)$; the matrices $S_\nu$ are defined as

\begin{equation}
S_\nu=\left(\begin{array}{cc}\Phi_1(\zeta_\nu)&-{\Phi^\ast _2(\zeta_\nu)}\\
\frac{1}{\zeta_\nu}\Phi_2(\zeta_\nu) &\zeta_\nu^\ast{\Phi^\ast _1(\zeta_\nu)}\end{array} \right)
\end{equation}
and are determined from Eqs.(\ref{syst}) using the seed solution Eq.(\ref{seed}).
In this way the $N$-soliton solution of Eqs.~(\ref{lat})  is put in the form
\begin{equation}
\beta_n=-\frac{i}{\sqrt{2\gamma}}{y^\ast _N}x_N,\,s_{n-1}=-\frac{1}{\sqrt{2\gamma}}\frac{{\cal Q}^\ast}{{\cal Q}}
\frac{{y^\ast _1}}{x_N}\end{equation}
\begin{equation} q^{-2(N_n+H_n)}= q^{-2H}\frac{x_N(n+1)}{x_N(n)},
\end{equation}
where $x_N(n,t)$ etc. is determined from Eqs.~(\ref{28})

 For the one-soliton case, $N=1$, we can choose the single point of the discrete spectrum $\zeta_0=e^{\gamma_0+i\alpha_0}$ (say)
and $\gamma_0<0$. We then find the formulae
\begin{equation}
\beta_n(t)=i\sqrt{\frac{2}{\gamma}}\sinh{(2\gamma_0)}
\frac{\exp{i(\phi(n,t)-\alpha_0)}}{\cosh{(\psi(n,t)-\gamma_0)}},\label{sola}
\end{equation}

 \begin{equation}
s_{n-1}(t)=-\sqrt{\frac{2}{\gamma}}\sinh{(2\gamma_0)}
\frac{\exp{i(\phi(n,t)+\alpha_0)}}{\cosh{(\psi(n,t)+\gamma_0)}}\label{solb}
\end{equation}

\begin{equation}
q^{2(N_n+H_n)}=
q^{2H}\frac{1-\tanh{(\psi(n,t)-\gamma_0)}\tanh{\vartheta_0}}{1-\tanh{(\psi(n,t)+\gamma_0)}\tanh{\vartheta_0}}.\label{solc}
\end{equation}

Here \be \phi(n,t)=t\cosh{(2\gamma_0)}\sin{(2\alpha_0)}-n\varrho_0+\phi_0,\label{par1}\ee

\be\psi(n,t)=t\sinh{(2\gamma_0)}\cos{(2\alpha_0)}-n\vartheta_0+\psi_0,\label{par2}\ee

\begin{eqnarray}
\vartheta_0&=&\frac{1}{2}\ln{\frac{\mbox{sinh}^2(\gamma_0-H\gamma)+
\mbox{sin}^2\alpha_0}{\mbox{sinh}^2(\gamma_0+H\gamma)+\mbox{sin}^2\alpha_0}}+2\gamma_0\label{par3}\\
\varrho_0&=&\mbox{arg}\frac{\sinh{(\gamma_0-H\gamma+i\alpha_0)}}{\sinh{(\gamma_0+H\gamma+i\alpha_0)}}+2\alpha_0,\label{par4}
\end{eqnarray}

$\phi_0$
and
$\psi_0$ are arbitrary real constants.

The formulae for various multisoliton solutions for the lattice system Eqs.~(\ref{lat}) are too complicated to be presented in detail
here. Since for the {\em lattice} these solutions  depend explicitly on the
deformation parameter $q=e^\gamma$, these $N$-soliton solutions ($N=1,2,\ldots$) are naturally thought of as {\em $q$-deformed
solitons}. In continuum limit Eq.~(\ref{22}), $q\to 1$ and $\gamma\to 0$ and $q$-deformation disappear.

\section{Conclusions and discussion}

 As was mentioned above in the case of the real (imaginary)
dynamical variables and in the sharp line limit the MB system
Eqs.~(\ref{mb}) is equivalent to the Sine-Gordon equation. The
same procedure is applicable to the LMB system which means that in
the case of the reduction to the real (imaginary) dynamical
variables the LMB system is in fact a new version of the lattice
Sine-Gordon equation. The dressing procedure described in this
paper can be extended to this case delivering the whole variety of
solutions of the (lattice) S-G equation (solitons, breathers
etc.).
 In so far as Eqs.~(\ref{sola})-(\ref{par4}) form
a $q$-deformed  soliton we can use this result to gain insight into the quantum case. One objective of the investigation of the
quantum MB system must be to find out, the precise nature of, and to calculate, the 'quantum soliton'
solutions. In Refs.~\cite{ryb1},\cite{ryb2} we introduced and solved exactly through the quantum inverse method (up to the
solutions of the Bethe equations)  a quantum version of the LMB system Eqs.~(\ref{lat}). Since this model provides a natural and
exactly solvable lattice regularization of the continuous limit quantum envelope MB  (or SIT) system (and recall that the quantum
Sine-Gordon can be embedded in this quantum MB) it is very useful for the construction  of the evolution operator and for
investigating the quantum dynamics of these continuous models which have direct physical meaning. It is known from a number of
quantum models~\cite{tyu,wad,resh} that a 'string solution' of the Bethe equations for the quantum model corresponds, in the limit of
a large number of collective excitations $M$, to the soliton solution of the classical counterpart of the exactly solvable quantum
system. A plausible conjecture which we will justify elsewhere is that the soliton solution Eq.~(\ref{sola}) for the 'electric field'
is given by the matrix element $\lim_{M\to\infty} \langle 0|
C(\lambda_1) C(\lambda_2)... C(\lambda_{M})\beta^{\dag}_nB(\lambda_1)B(\lambda_1)...B(\lambda_{M-1})|0\rangle$, where
$B(\lambda)$ is a creation operator for a quasiparticle and $C(\lambda)=B^{\dag}(\lambda)$ is an annihilation operator.
The rapidities $\lbrace\lambda_l \rbrace_{l=1}^M$ are roots of the Bethe equations
$
e^{2i\gamma
M\lambda_n}\frac{\mbox{sin}^M\gamma(\lambda_l-iS)}{\mbox{sin}^M
\gamma(\lambda_l+iS)}
=\prod_{j=1}^N\frac{\sin{\gamma(\lambda_l-\lambda_j-i)}}{\sin{\gamma(\lambda_l-\lambda_j-i)}}
$.The operator $\beta_n^{\dag}$ is the electric field operator
which satisfies the $q$-deformed $q$-boson algebra analogous to
the algebra Eq.~(\ref{bos}). In Ref.\cite{resh} it was shown that
the creation operator $B(\lambda)$ plays the role of the quantum
counterpart of a Blaschke multiplier which builds the classical
soliton solution~\cite{fad}. The dressing operator $F(\zeta)$
Eq.~(\ref{backl1}) up to certain modifications has the same
meaning. This indicates very well how the experience obtained in
the analysis of the $c$-number system reported in this paper can
be used in understanding of the quantum case. In practice this
experience helps us to trace out the formation of the classical
optical soliton from a large number of quantum collective
excitations, a physical problem of considerable interest\cite{17}.

\section*{Acknowledgements}

One of us (RKB) is grateful to Dr. P.J. Caudrey for helpful
discussions of the soliton solutions reported in this paper, GV
was partly supported by a Russian Federation research grant RFBR
No 98-01-01063


\begin{thebibliography}{99}
\bibitem{17} Quantum features are detected in optical solitons in
fibres in~\cite{18},\cite{19} but 'quantum solitons' as in
e.g.~\cite{20} may not necessarily mean the true solitons
considered in this paper.
\end{thebibliography}
\end{document}